\documentclass[preprint,preprintnumbers,aps,12pt,superscriptaddress,nofootinbib,tightenlines,floatfix]{revtex4-1}

\usepackage{amsmath,amssymb}
\usepackage{graphicx}
\usepackage{bm}
\usepackage{comment}
\usepackage{color}

\newcommand{\gsim}{\raisebox{-0.7ex}{$\stackrel{\textstyle >}{\sim}$ }}
\newcommand{\lsim}{\raisebox{-0.7ex}{$\stackrel{\textstyle <}{\sim}$ }}
\usepackage[T1]{fontenc}
\usepackage[latin9]{inputenc}
\usepackage{graphicx}
\usepackage{esint}
\usepackage{hyperref}
\usepackage{atbegshi}
\usepackage{lipsum}

\def\Dslash{D\hskip-0.65em /}

\def\SW{Sheikholeslami-Wohlert}
\def\lattspace{b}

\newcount\hour \newcount\hourminute \newcount\minute 
\hour=\time \divide \hour by 60
\hourminute=\hour \multiply \hourminute by 60
\minute=\time \advance \minute by -\hourminute
\newcommand{\mydate}{\ \today \ - \number\hour :\number\minute}

\begin{document}

\dimen\footins=8\baselineskip\relax

\preprint{
\vbox{ 
\hbox{NT@UW-12-14}
\hbox{INT-PUB-12-046}
}}

\title{\textbf{Constraints on the  Universe as a Numerical Simulation}}

\author{Silas R.~Beane}
\email{beane@hiskp.uni-bonn.de.  On leave from the University of New Hampshire.}
\affiliation{Institute for Nuclear Theory, Box 351550, Seattle, WA 98195-1550, USA}
\affiliation{Helmholtz-Institut f\"ur Strahlen- und Kernphysik (Theorie),
Universit\"at Bonn, D-53115 Bonn, Germany}
\author{Zohreh Davoudi}
\email{davoudi@uw.edu}
\affiliation{Department of Physics,
  University of Washington, Box 351560, Seattle, WA 98195, USA}

\author{Martin J.~Savage}
\email{savage@phys.washington.edu}
\affiliation{Department of Physics,
  University of Washington, Box 351560, Seattle, WA 98195, USA}

\date{\mydate}

\begin{abstract}
  Observable consequences of the hypothesis that the observed universe
  is a numerical simulation performed on a cubic space-time lattice or grid
  are explored. 
The simulation scenario is first motivated by
  extrapolating current trends in computational resource requirements
  for lattice QCD into the future.  
Using the historical development
  of lattice gauge theory technology as a guide, we assume that our
  universe is an early numerical simulation with unimproved Wilson
  fermion discretization and investigate potentially-observable
  consequences.  
Among the observables that are considered are the muon $g-2$ and the current
differences between determinations of $\alpha$, but
the most stringent bound on the inverse lattice spacing 
of the universe, $b^{-1}\gsim 10^{11}~{\rm GeV}$,
is derived  from the high-energy cut off of the cosmic ray spectrum.
The numerical simulation scenario
  could reveal itself in the distributions of the highest energy
  cosmic rays exhibiting a degree of
rotational symmetry breaking that reflects the structure of the underlying lattice.
\end{abstract}
\maketitle
\vfill\eject

\section{Introduction}
\label{sec:Intro}

\noindent Extrapolations to the distant futurity of trends in the
growth of high-performance computing (HPC) have led philosophers to
question ---in a logically compelling way--- whether the universe that
we currently inhabit is a numerical simulation performed by our
distant descendants~\cite{Bostrom2003}.  With the current developments
in HPC and in algorithms it is now possible to simulate Quantum
Chromodynamics (QCD), the fundamental force in nature that gives rise
to the strong nuclear force among protons and neutrons, and to nuclei
and their interactions.  These simulations are currently performed in
femto-sized universes where the space-time continuum is replaced by a
lattice, whose spatial and temporal sizes are of the order of several
femto-meters or fermis ($1~{\rm fm}=10^{-15}~{\rm m}$), and whose
lattice spacings (discretization or pixelation) are fractions of
fermis~\footnote{Surprisingly,
  while QCD and the electromagnetic force are currently being
  calculated on the lattice, the difficulties in simulating the weak
  nuclear force and gravity on a lattice have so far proved
  insurmountable.}.  This endeavor, generically referred to as lattice
gauge theory, or more specifically lattice QCD, is currently leading
to new insights into the nature of matter~\footnote{ See Refs.~\protect\cite{Kronfeld:2012ym,Fodor:2012gf} for
  recent reviews of the progress in using lattice gauge theory to
  calculate the properties of matter.  }.  Within the next decade,
with the anticipated deployment of exascale computing resources, it is
expected that the nuclear forces will be determined from QCD, refining
and extending their current determinations from experiment, enabling
predictions for processes in extreme environments, or of exotic forms
of matter, not accessible to laboratory experiments.  Given the
significant resources invested in determining the quantum fluctuations
of the fundamental fields which permeate our universe, and in
calculating nuclei from first principles (for recent works, see
Refs.~\cite{Beane:2012vq,Yamazaki:2012hi,Aoki:2012tk}), it stands to
reason that future simulation efforts will continue to extend to
ever-smaller pixelations and ever-larger volumes of space-time, from
the femto-scale to the atomic scale, and ultimately to macroscopic
scales. If there are sufficient HPC resources available, then future
scientists will likely make the effort to perform complete simulations
of molecules, cells, humans and even beyond.  Therefore, there is a
sense in which lattice QCD may be viewed as the nascent science of
universe simulation, and, as will be argued in the next paragraph,
very basic extrapolation of current lattice QCD resource trends into
the future suggest that experimental searches for evidence that our
universe is, in fact, a simulation are both interesting and logical.

There is an extensive literature which explores various aspects of our
universe as a simulation, from philosophical
discussions~\cite{Bostrom2003}, to considerations of the limits of
computation within our own universe~\cite{Lloyd:1999}, to the
inclusion of gravity and the standard model of particle physics into a
quantum computation~\cite{Lloyd:2005js}, and to the notion of our
universe as a cellular
automaton~\cite{Zuse1969rr,Fredkin:1990,wolfram02,'tHooft:2012uy}. There have
also been extensive connections made between fundamental aspects of
computation and physics, for example, the translation of the
Church-Turing principle~\cite{Church:1936,Turing:1936} into the
language of physicists by Deutsch~\cite{Deutsch:1985}. Finally, the
observational consequences due to limitations in accuracy or flaws in
a simulation have been considered~\cite{Barrow:2008}.  In this work,
we take a pedestrian approach to the possibility that our universe is
a simulation, by assuming that a classical computer (i.e. the
classical limit of a quantum computer) is used to simulate the quantum
universe (and its classical limit), as is done today on a very small
scale, and ask if there are any signatures of this scenario that might
be experimentally detectable.  Further, we do not consider the
implications of, and constraints upon, the underlying information, and
its movement, that are required to perform such extensive simulations.
It is the case that the method of simulation, the algorithms, and the
hardware that are used in future simulations are unknown, but it is
conceivable that some of the ingredients used in present day
simulations of quantum fields remain in use, or are used in other
universes, and so we focus on one aspect only: the possibility that
the simulations of the future employ an underlying cubic lattice
structure.

\begin{figure}[!t]
  \centering
     \includegraphics[scale=0.40]{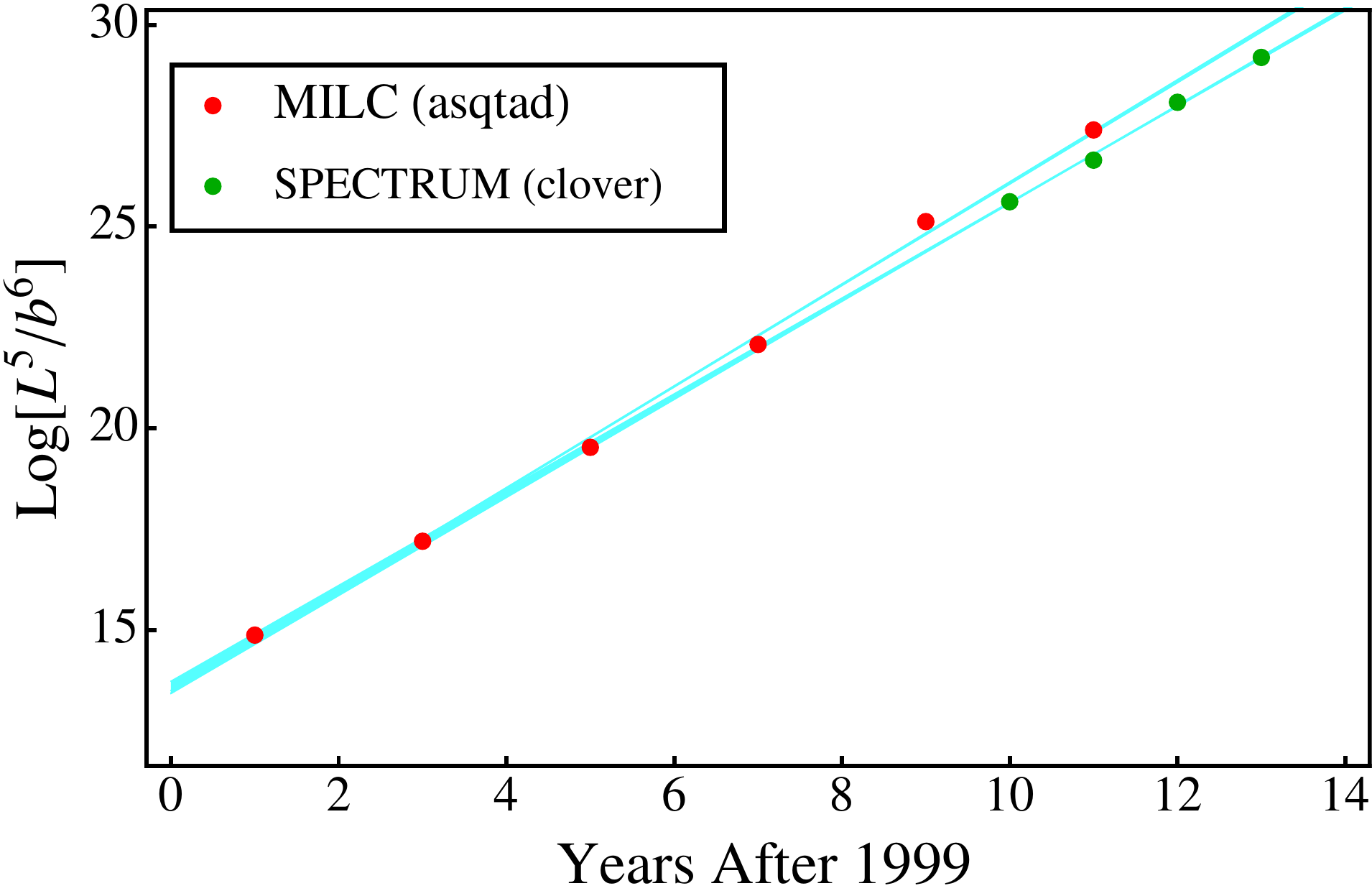} \ \ \ \ \includegraphics[scale=0.41]{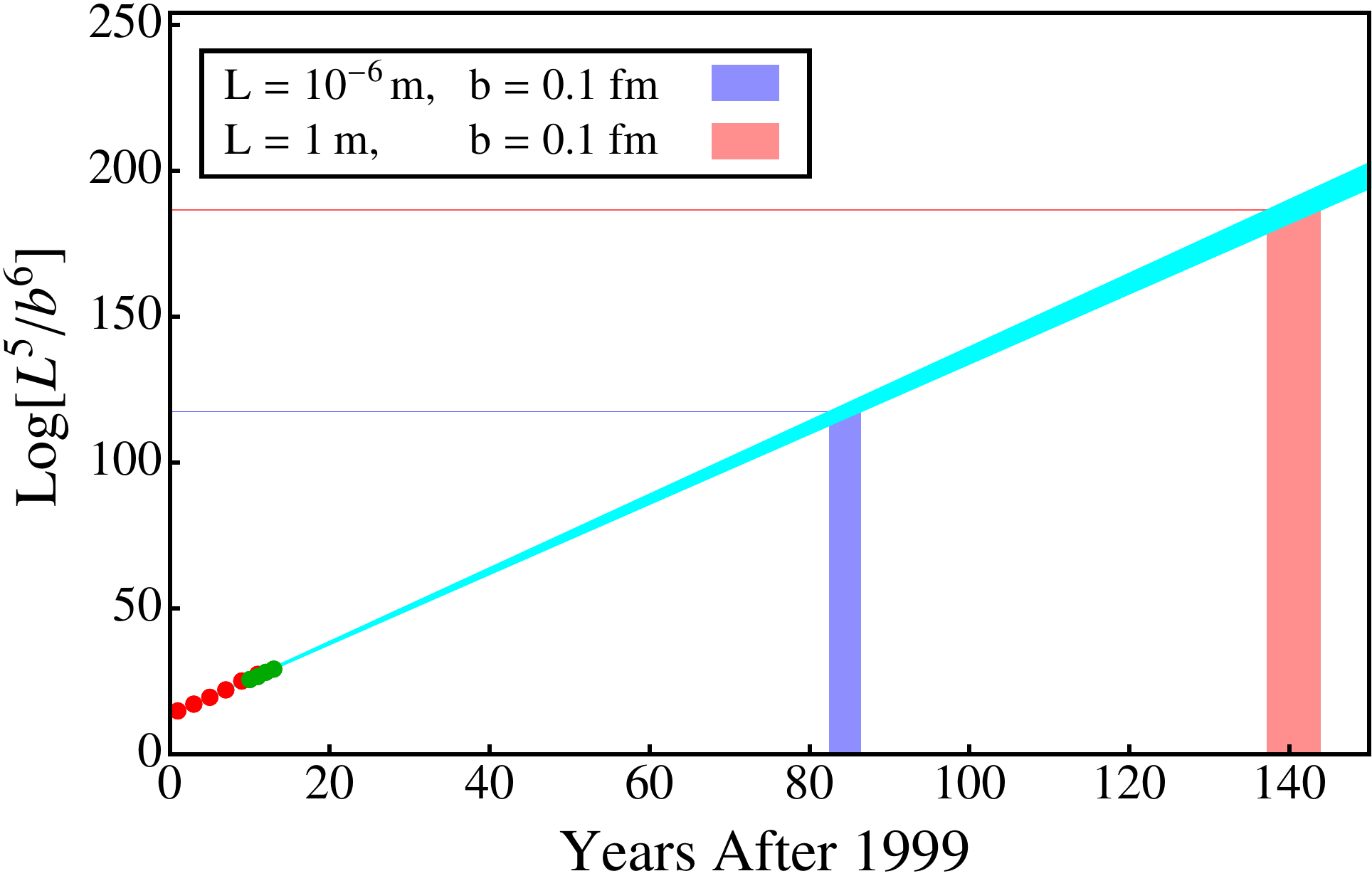}
     \caption{Left panel: linear fit to the logarithm of the CRRs of
       MILC asqtad and the SPECTRUM anisotropic lattice ensemble generations. 
Right panel:
       extrapolation of the fit curves into the future, as discussed in
       the text. 
The blue (red) horizontal line corresponds to lattice sizes of one micron (meter), 
and vertical bands show the corresponding extrapolated years beyond 1999 for
the lattice generation programs.
}
  \label{fig:MILCextrap}
\end{figure}

In contrast with Moore's law, which is a statement about the
exponential growth of raw computing power in time, it is interesting
to consider the historical growth of measures of the computational
resource requirements (CRRs) of lattice QCD calculations, and
extrapolations of this trend to the future. In order to do so, we
consider two lattice generation programs: the MILC asqtad 
program~\cite{MILC}, which over a twelve year span generated ensembles
of lattice QCD gauge configurations, using the
Kogut-Susskind~\cite{Kogut:1974ag} (staggered) discretization of the
quark fields, with lattice spacings, $b$, ranging from $0.18$ to
$0.045~$fm, and lattice sizes (spatial extents), $L$, ranging from
$2.5$ to $5.8~$fm, and the on-going anisotropic program carried out by
the SPECTRUM collaboration~\cite{SPECTRUM}, using the
clover-Wilson~\cite{Wilson:1974sk,Sheikholeslami} discretization of
the quark fields, which has generated lattice ensembles at $b\sim
0.1~$fm, with $L$ ranging from $2.4$ to $4.9~$fm~\cite{Lin:2008pr}.
At fixed quark masses, the CRR of a lattice ensemble generation (in
units of petaFLOP-years) scales roughly as the dimensionless number
$\lambda_{QCD}L^5/b^6$, where $\lambda_{QCD}\equiv 1~$fm is a typical
QCD distance scale.  In fig.~\ref{fig:MILCextrap} (left panel), the
CRRs are presented on a logarithmic scale, where year one corresponds
to 1999, when MILC initiated its asqtad program of $2+1$-flavor
ensemble generation. The bands are linear fits to the data.  While the
CRR curves in some sense track Moore's law, they are more than a
statement about increasing FLOPS. 
Since lattice QCD simulations include the quantum fluctuations of the vacuum and the effects of the strong
nuclear force, the CRR curve is a statement about simulating universes
with realistic fundamental forces.  The extrapolations of the CRR
trends into the future are shown in the right panel of
fig.~\ref{fig:MILCextrap}.
The blue (red) horizontal line corresponds to a lattice of the size of a
micro-meter (meter), a typical length scale of a cell (human), and at a
lattice spacing of $0.1~$fm.  There are, of course, many caveats to
this extrapolation.  Foremost among them is the assumption that an
effective Moore's Law will continue into the future, which requires
technological and algorithmic developments to continue as they have
for the past 40 years.  Related to this is the possible existence of
the technological singularity~\cite{Vinge,Kurzweil:2006}, which could
alter the curve in unpredictable ways.  And, of course, human
extinction would terminate the exponential growth~\cite{Bostrom2003}.
However, barring such discontinuities in the curve, these estimates
are likely to be conservative as they correspond to full simulations
with the fundamental forces of nature. With finite resources at their
disposal, our descendants will likely make use of effective theory
methods, as is done today, to simulate every-increasing complexity,
by, for instance, using meshes that adapt to the relevant physical
length scales, or by using fluid dynamics to determine the behavior of
fluids, which are constrained to rigorously reproduce the fundamental
laws of nature. Nevertheless, one should keep in mind that the
CRR curve is based on lattice QCD ensemble generation and therefore
is indicative of the ability to simulate the quantum fluctuations
associated with the fundamental forces of nature at a given lattice
spacing and size. The cost to perform the measurements
that would have to be done in the background of these fluctuations in
order to simulate ---for instance--- a cell could, in principle, 
lie on a significantly steeper curve.

We should comment on the simulation scenario in the context of ongoing
attempts to discover the theoretical structure that underlies the
Standard Model of particle physics, and the expectation of the
unification of the forces of nature at very short distances. There has
not been much interest in the notion of an underlying lattice
structure of space-time for several reasons.  Primary among them is
that in Minkowski space, a non-vanishing spatial lattice spacing
generically breaks space-time symmetries in such a way that there are
dimension-four Lorentz breaking operators in the Standard Model,
requiring a large number of fine-tunings to restore Lorentz invariance
to experimentally verified levels~\cite{Coleman:1998ti}. The fear is
that even though Lorentz violating dimension four operators can be
tuned away at tree-level, radiative corrections will induce them back
at the quantum level as is discussed in Refs. \cite{Gagnon:2004xh,
  Collins:2004bp}. 
This is not an issue if one assumes the
simulation scenario for the same reason that it is not an issue when
one performs a lattice QCD calculation~\footnote{Current lattice QCD
  simulations are performed in Euclidean space, where the underlying
  hyper-cubic symmetry protects Lorentz invariance breaking in dimension four
  operators. However, Hamiltonian lattice formulations, which are
  currently too costly to be practical, are also possible.}.  
The underlying space-time symmetries respected by the lattice action will necessarily
be preserved at the quantum level.  In addition, the notion of a
simulated universe is sharply at odds with the reductionist prejudice
in particle physics which suggests the unification of forces with a
simple and beautiful predictive mathematical description at very short
distances. However, the discovery of the string
landscape~\cite{Kachru:2003aw,Susskind:2003kw}, and the current
inability of string theory to provide a useful predictive framework
which would post-dict the fundamental parameters of the Standard
Model, provides {\it the simulators} (future string theorists?) with a
purpose: to systematically explore the landscape of vacua through
numerical simulation. If it is indeed the case that the fundamental
equations of nature allow on the order of $10^{500}$
solutions~\cite{Douglas:2003um}, then perhaps the most profound quest
that can be undertaken by a sentient being is the exploration of the
landscape through universe simulation.   In some weak sense,
this exploration is already underway with current investigations of a
class of confining beyond-the-Standard-Model (BSM) theories, where
there is only minimal experimental guidance at present (for one recent
example, see Ref.~\cite{Appelquist:2012nz}).  
Finally, one may be
tempted to view lattice gauge theory as a primitive numerical tool,
and that the simulator should be expected to have more efficient ways
of simulating reality. However, one should keep in mind that the only
known way to define QCD as a consistent quantum field theory is in the
context of lattice QCD, which suggests a fundamental role for the
lattice formulation of gauge theory.

Physicists, in contrast with philosophers, are interested in
determining observable consequences of the hypothesis that we are a
simulation~\footnote{
There are a number of peculiar observations that
  could be attributed to our universe being a simulation, but that 
  cannot be tested at present. For instance, it could be that the observed
  non-vanishing value of the cosmological constant is simply a
  rounding error resulting from the number zero being entered into a simulation
  program with insufficient precision.}~\footnote{
Hsu and Zee~\cite{Hsu:2005nn} have suggested that the CMB provides an
opportunity for a potential creator/simulator of our universe to
communicate with the created/simulated without further intervention in the
evolution of the universe.  If, in fact, it is determined that
observables in our universe are consistent with those that would
result from a numerical simulation, then the Hsu-Zee scenario becomes
a more likely possibility.  Further, it would then become interesting
to consider the possibility of communicating with the simulator, or
even more interestingly, manipulating or controlling the simulation
itself.
}.  
In lattice QCD, space-time is
replaced by a finite hyper-cubic grid of points over which the fields
are defined, and the (now) finite-dimensional quantum mechanical path
integral is evaluated. The grid breaks Lorentz symmetry (and hence
rotational symmetry), and its effects have been defined within the
context of a low-energy effective field theory (EFT), the Symanzik
action, when the lattice spacing is small compared with any physical
length scales in the
problem~\cite{SymanzikI,SymanzikII}~\footnote{The finite volume of the
  hyper-cubic grid also breaks Lorentz symmetry.  
A recent analysis of the CMB suggests that universe
has a compact topology, consistent with two compactified spatial 
dimensions and with a greater than
$4\sigma$ deviation from three uncompactified spatial dimensions~\cite{Aslanyan:2011zp}.
}. 
The lattice action can be modified to 
systematically improve calculations of observables, by adding
irrelevant operators with coefficients that can be determined
nonperturbatively.  For instance, the Wilson action can be ${\cal
  O}(b)$-improved by including the
{\SW} term~\cite{Sheikholeslami}.  Given this low-energy
description, we would like to investigate the hypothesis that we are a
simulation with the assumption that the development of simulations of
the universe in some sense parallels the development of lattice QCD
calculations. That is, early simulations use the computationally ``cheapest''
discretizations with no improvement.  
In particular, we will assume
that the simulation of our universe is done on a hyper-cubic grid~\footnote{The
concept of the universe consisting of fields defined on nodes, and interactions
propagating along the links  between the nodes,
separated by distances of order the Planck length, 
has been considered previously,
e.g. see Ref.~\cite{Jizba:2009qf}.}
and, as a starting point, 
we will assume that {\it the simulator} is
using an unimproved Wilson action, that produces ${\cal O}(b)$
artifacts of the form of the {\SW} 
operator in the low-energy theory~\footnote{It has been recently pointed
  out that the domain-wall formulation of lattice fermions
provides  a mechanism by which the number of generations of fundamental 
particles is tied to the form of the dispersion relation~\cite{Kaplan:2011vz}.
Space-time would then be a topological insulator.
}.

In section~\ref{sec:unimproved}, the simple scenario of
an unimproved Wilson action is introduced. In section~\ref{sec:RIM}, by looking at the
rotationally-invariant dimension-five operator arising from this action, the bounds on
the lattice spacing are extracted from the current experimental
determinations, and theoretical calculations, of 
$g-2$ of the electron and muon, 
and from the fine-structure constant, $\alpha$, determined by  the Rydberg
constant. Section~\ref{sec:RSB} considers the simplest effects of
Lorentz symmetry  breaking operators that first appear at ${\cal
  O}(\lattspace^2)$, and modifications to the energy-momentum
relation. 
Constraints on the energy-momentum relation due to cosmic
ray events are found to provide the most stringent bound on
$\lattspace$. We conclude in section~\ref{sec:conc}.

\section{Unimproved Wilson Simulation of the Universe}
\label{sec:unimproved}
\noindent

\noindent The simplest gauge invariant action of fermions which does not
contain  doublers is the Wilson action,
\begin{eqnarray}
S^{(W)}  & = & \lattspace^4 \sum_{x} {\cal L}^{(W)}(x)
 = \lattspace^4 
\left( m + {4\over \lattspace} \right)\ \sum_{x}\ \overline{\psi}(x)\psi(x)
\nonumber\\
 & + & 
{\lattspace^{3}\over 2} \sum_{x}\ \overline{\psi}(x)
\left[\, \left(\gamma_\mu-1\right)\ U_\mu(x)\ \psi(x+\lattspace\hat\mu)
 - 
\left(\gamma_\mu+1\right)\ U^\dagger_\mu(x-\lattspace\hat\mu)\ \psi(x-\lattspace\hat\mu)
\, \right],
\label{eq:wilson}
\end{eqnarray}
which describes a fermion, $\psi$, of
mass $m$ interacting with a gauge field, $A_\mu(x)$, through the gauge link,
\begin{eqnarray}
U_\mu(x)
& = & 
\exp\ \left( ig\int_x^{x+\lattspace\hat\mu}\ dz A_\mu(z) \ \right),
\label{eq:link}
\end{eqnarray}
where $\hat\mu$ is a unit vector in the $\mu$-direction, and $g$ is the coupling constant of the theory.
Expanding the Lagrangian density, ${\cal L}^{(W)}$, in the lattice spacing (that
is small compared with the physical length scales), 
and performing a field redefinition~\cite{Luscher:1984xn},
it can be shown that the Lagrangian density takes the form
\begin{eqnarray}
{\cal L}^{(W)} 
& = & 
\overline{\psi}\Dslash\psi\ +\ \tilde m \overline{\psi}\psi\
\ +\ {\cal C}_p{g \lattspace\over 4} \overline{\psi}\sigma_{\mu\nu}G^{\mu\nu} \psi
\ +\ {\cal O}(\lattspace^2),
\label{eq:effd}
\end{eqnarray}
where $G_{\mu\nu}=-i[\,D_\mu\,,\,D_\nu\,]/g$ is the field strength
tensor and $D_{\mu}$ is the covariant derivative. $\tilde m$ is
a redefined mass which contains ${\cal O}(\lattspace)$ lattice spacing
artifacts (that can  be tuned away). The coefficient of the Pauli term
$\overline{\psi}\sigma_{\mu\nu}G^{\mu\nu} \psi$ is fixed at tree
level, ${\cal C}_p=1+{\cal O}(\alpha)$, where
$\alpha = g^2/(4\pi)$.  It is worth noting that as is
usual in lattice QCD calculations, the lattice action can be ${\cal
  O}(\lattspace)$ improved by adding a term of the form $\delta{\cal
  L}^{(W)} = C_{sw}{g \lattspace\over 4}
\overline{\psi}\sigma_{\mu\nu}G^{\mu\nu} \psi
$
to the Lagrangian with $C_{sw}=-{\cal C}_p +{\cal O}(\alpha)$. This is the
so-called {\SW}  term. 
Of course there is no reason to assume that \textit{the simulator} had to have
performed such an 
improvement in simulating the universe.

\section{Rotationally Invariant Modifications}
\label{sec:RIM}
\noindent
Lorentz symmetry is recovered in lattice calculations as the lattice
spacing vanishes when compared with the scales of the system.  
It is useful to consider
contributions to observables from a non-zero lattice spacing that are
Lorentz invariant and consequently rotationally invariant, 
and those that are not. 
While the former type of modifications could arise from many different BSM
scenarios, the latter, 
particularly modifications that
exhibit cubic symmetry, would be suggestive of a structure consistent
with an underlying discretization of space-time.

\subsubsection{QED Fine Structure Constant and the Anomalous Magnetic Moment}
\noindent
For our present purposes, we will assume that Quantum Electrodynamics
(QED) is simulated with this unimproved action,
eq. (\ref{eq:wilson}). The ${\cal O}(\lattspace)$ contribution to the
lattice action induces an additional contribution to the fermion
magnetic moments.  Specifically, the Lagrange density that describes
electromagnetic interactions is given by eq. (\ref{eq:effd}), where
the interaction with an external magnetic field $\mathbf{B}$ is
described through the covariant derivative $D_\mu = \partial_\mu + i e
{\hat Q} A_\mu$ with $e>0$ and the electromagnetic charge operator
${\hat Q}$, and where the vector potential satisfies
$\nabla\times\mathbf{A}=\mathbf{B}$.  
The interaction Hamiltonian density in Minkowski-space is
given by
\begin{eqnarray}
{\cal H}_{int} 
& = & 
{e\over 2 m} \overline{\psi} A_\mu (i\overrightarrow \partial^\mu -
i\overleftarrow\partial^\mu) {\hat Q} \psi 
\ +\ {{\hat Q} e\over 4 m} \overline{\psi}\sigma_{\mu\nu}F^{\mu\nu} \psi
\ +\ {\cal C}_p{{\hat Q} e \lattspace\over 4} \overline{\psi}\sigma_{\mu\nu}F^{\mu\nu} \psi
\ +\ ...
\ .
\label{eq:EMSWb}
\end{eqnarray}
where $F_{\mu\nu}=\partial_{\mu}A_{\nu}-\partial_{\nu}A_{\mu}$ 
is the electromagnetic field strength tensor, 
and the ellipses denote terms suppressed by additional powers of $\lattspace$.
By performing a non-relativistic reduction, the 
last two terms in eq. \ref{eq:EMSWb} give rise to 
$\mathcal{H}_{int,mag}=-{\bm\mu}\cdot {\bm B}$, 
where the  electron magnetic moment $\bm{\mu}$ is given by
\begin{eqnarray}
{\bm\mu}
& = & 
{{\hat Q} e\over 2 m} \left( g + 2 m \lattspace \ {\cal C}_p + ...\right)\ {\bm S}
\ =\ 
g(b) \ {{\hat Q} e\over 2 m}\ {\bm S}
\ ,
\label{eq:mueint}
\end{eqnarray}
where $g$ is the usual fermion g-factor and ${\bm S}$ is its spin.
Note that the lattice spacing contribution to the magnetic moment is
enhanced relative to the Dirac contribution by one power of the particle mass.

For the electron, the effective $g$-factor has an expansion at finite lattice spacing of
\begin{eqnarray}
\frac{g^{(e)}(b)}{2}
& = & 
1 + C_2 \left({\alpha\over\pi}\right)
 + C_4 \left({\alpha\over\pi}\right)^2
 + C_6 \left({\alpha\over\pi}\right)^3
 + C_8 \left({\alpha\over\pi}\right)^4
 + C_{10} \left({\alpha\over\pi}\right)^{5}
\nonumber\\
&& 
\ +\ a_{\rm hadrons}\ +\ a_{\mu,\tau}\ +\ a_{\rm weak}
\ +\
m_e \lattspace  \ {\cal C}_p
\ +\ ...
\ \ \ ,
\label{eq:ge}
\end{eqnarray}
where the coefficients $C_i$, in general, depend upon the ratio of lepton
masses.  
The calculation by Schwinger provides the
leading coefficient of  $C_2={1\over 2}$.
The experimental value of 
$g^{(e)}_{\rm expt}/2 = 
1.001\ 159\ 652\ 180\ 73(28)$
gives rise to the best determination of the fine structure constant
$\alpha$ (at $\lattspace=0$)~\cite{Mohr:2012tt}.  However, when the lattice
spacing is non-zero,
the extracted value of $\alpha$  becomes a function of $\lattspace$,
\begin{eqnarray}
\alpha(\lattspace) & = & 
\alpha(0)\ -\ 2\pi m_e \lattspace \ {\cal C}_p\ +\  \mathcal{O} \left(\alpha ^{2}b \right)
\ \ \ ,
\label{eq:alphaexp}
\end{eqnarray}
where $\alpha(0)^{-1} = 137.035\ 999\ 084(51)$ is determined from the
experimental value of electron g-factor as quoted above.  With one
experimental constraint and two parameters to determine, $\alpha$ and
$b$, unique values for these quantities cannot be established, and an
orthogonal constraint is required. One can look at the muon $g-2$
which has a similar QED expansion to that of the electron, including
the contribution from the non-zero lattice spacing,
\begin{eqnarray}
\frac{g^{(\mu)}(b)}{2}
& = & 
1 + C^{(\mu)}_2 \left({\alpha\over\pi}\right)
 + C^{(\mu)}_4 \left({\alpha\over\pi}\right)^2
 + C^{(\mu)}_6 \left({\alpha\over\pi}\right)^3
 + C^{(\mu)}_8 \left({\alpha\over\pi}\right)^4
 + C^{(\mu)}_{10} \left({\alpha\over\pi}\right)^{5}
\nonumber\\
&& 
\ +\ a^{(\mu)}_{\rm hadrons}\ +\ a^{(\mu)}_{e,\tau}\ +\ a^{(\mu)}_{\rm weak}
\ +\
m_\mu \lattspace  \ {\cal C}_p
\ +\ ...
\ \ \ .
\label{eq:gmuon}
\end{eqnarray}
Inserting the 
electron  $g-2$ (at finite lattice spacing)
gives
\begin{eqnarray}
\frac{g^{(\mu)}(b)}{2}
& = & 
\frac{g^{(\mu)}(0)}{2}
\ +\
(m_\mu-m_e) \lattspace\  {\cal C}_p \ + \  \mathcal{O} \left(\alpha ^{2}b \right)
\ \ \ .
\label{eq:gmuonexp}
\end{eqnarray}
Given that the standard model calculation of $g^{(\mu)}(0)$
is consistent with the experimental value, with a $\sim 3.6\sigma$
deviation, one can put a limit on $ \lattspace$ from the difference
and uncertainty in theoretical and experimental values of $g^{(\mu)}$,
$g^{(\mu)}_{\rm expt}/2 = 1.001\ 165\ 920\ 89(54)(33)$ and
$g^{(\mu)}_{\rm theory}/2 = 1.001\ 165\ 918\
02(2)(42)(26)$~\cite{Mohr:2012tt}.
Attributing this difference to a finite lattice spacing,
these values give rise to 
\begin{eqnarray}
\lattspace^{-1} & = & 
\left( 3.6\pm 1.1 \right) \times 10^{7}~{\rm GeV}
\ \ \ ,
\label{eq:binvfixmuon}
\end{eqnarray}
which provides an approximate upper  bound on  the lattice spacing.

\subsubsection{The Rydberg Constant and $\alpha$}
\label{subsec:Rydberg-alpha}
\noindent
Another limit can be placed on the lattice spacing from differences
between the value of $\alpha$ extracted from the electron $g-2$ and
from the Rydberg constant, $R_\infty$.  The latter extraction, as
discussed in Ref.~\cite{Mohr:2012tt}, is rather complicated, with the
value of the $R_\infty$ obtained from a $\chi^2$-minimization fit 
involving
the experimentally determined energy-level splittings.  However, to
recover the constraints on the Dirac energy-eigenvalues (which then
lead to $R_\infty$), theoretical corrections must be first removed from
the experimental values.  
To begin with, one can obtain an estimate
for the limit on $\lattspace$ by considering the differences between
$\alpha$'s obtained from various methods assuming that the only
contributions are from QED and the lattice spacing.  
Given that 
it is the reduced mass ($\mu\sim m_e$) that will 
compensate the lattice spacing in these QED
determinations (for an atom at rest in the lattice frame), one can write
\begin{eqnarray}
\delta\alpha & = & 2 \pi m_e \lattspace\ \mathcal{\tilde C}_{p}
\ \ \ ,
\label{eq:delalpha}
\end{eqnarray}
where $\mathcal{\tilde C}_{p}$ is a number ${\cal O}(1)$ by
naive dimensional analysis, and is a combination of the contributions
from the two independent extractions of $\alpha$.  There is no reason
to expect complete cancellation between the contributions from two
different extractions. In fact, it is straightforward to show that the
${\cal O}(b)$ contribution to the value of $\alpha$ determined from
the Rydberg constant is suppressed by $\alpha^{4}m_{e}^{2}$, and
therefore the above assumption is robust.  In addition to the electron
$g-2$ determination of fine structure constant as quoted above, the
next precise determination of $\alpha$ comes form the atomic recoil
experiments,
$\alpha^{-1} = 137.035\ 999\ 049(90)$~\footnote{Extracted from a
  $^{87} Rb$ recoil experiment~\cite{Bouchendira}.}
~\cite{Mohr:2012tt}, given 
an a priori determined value of the Rydberg
constant.
This gives rise to a difference of $|\delta\alpha| = \left(1.86\pm
  5.51\right)\times 10^{-12}$ 
between two extractions,
which translates into 
\begin{eqnarray}
\lattspace & = & 
| \left( -0.6\pm 1.7\right)\times 10^{-9} |~{\rm GeV}^{-1}
\ \ \ .
\label{eq:bfixAtomes}
\end{eqnarray}
As this result is consistent with zero, the 
$1\sigma$ values of 
the lattice spacing give rise to  a limit of
\begin{eqnarray}
\lattspace^{-1} & \gsim & 4\times 10^8~{\rm GeV}
\ \ \ ,
\label{eq:MfixAtomes}
\end{eqnarray}
which is seen to be an order of magnitude more precise than that
arising from the muon $g-2$.

For more sophisticated simulations in which chiral symmetry is preserved by the lattice
discretization, the coefficient $ {\cal C}_p$ will vanish or will be exponentially small.
As a result, the bound on the lattice spacing derived from the muon $g-2$ 
and from the differences between determinations of $\alpha$
will be significantly weaker.
In these analyses, we have worked with QED only, and have not included the full
electroweak interactions as chiral gauge theories have not yet been
successfully latticized.  Consequently, these constraints are to be considered
estimates only, and a more complete analysis needs to be performed when chiral
gauge theories can be simulated.

\section{Rotational Symmetry Breaking}
\label{sec:RSB}
\noindent
While there are more conventional scenarios for BSM physics
that generate deviations in $g-2$ from
the standard model prediction, or differences between independent
determinations of $\alpha$, the breaking of rotational symmetry would
be a solid indicator of an underlying space-time grid, although not the
only one.  As has been extensively discussed in the literature, another
scenario that gives rise to rotational invariance violation involves
the introduction of an external background with a preferred
direction. Such a preferred direction can be defined via a fixed
vector, $u_{\mu}$~\cite{Colladay}. 
The effective low-energy Lagrangian of
such a theory contains Lorentz covariant higher dimension operators
with a coupling to this background vector, and breaks both parity and
Lorentz invariance~\cite{Myers:2003fd}.  Dimension three, four and
five operators, however, are shown to be severely constrained by
experiment, and such contributions in the low-energy
action (up to dimension five) have been ruled out~\cite{Colladay,
  Coleman:1998ti, Carroll, Laurent:2011he}.

\subsubsection{Atomic Level Splittings}

\noindent 
At ${\cal O}(\lattspace^2)$ in the lattice spacing expansion of the
Wilson action, that is relevant to describing low-energy processes,
there is a rotational-symmetry breaking operator that is consistent
with the lattice hyper-cubic symmetry,
\begin{eqnarray}
{\cal L}^{RV} & = & 
C^{RV}\ {\lattspace^2\over 6}\ \sum_{\mu=1}^4\  \overline{\psi} \ \gamma_\mu D_\mu D_\mu D_\mu\
\psi
\ \ \ ,
\label{eq:RVopb}
\end{eqnarray}
where the tree-level value of $C^{RV}=1$. In taking matrix elements
of this operator in the Hydrogen atom, where the binding energy is
suppressed by a factor of $\alpha$ compared with the typical momentum, the
dominant contribution is from the spatial components. As each spatial momentum scales as $m_e \alpha$, in the non-relativistic limit, 
shifts in the energy levels are expected to be of order 
\begin{eqnarray}
\delta E & \sim & 
C^{RV}\ \alpha^4 m_e^3 \lattspace^2  \ .
\label{eq:J52energies}
\end{eqnarray}
To understand the size of energy splittings, a lattice spacing of
$\lattspace^{-1} = 10^8~{\rm GeV}$ gives an energy shift of order
$\delta E\sim 10^{-26}~{\rm eV}$, including for the splittings
between substates in given irreducible representations of SO(3) 
with angular momentum $J\geq 2$.
This magnitude of energy shifts and
splittings is presently unobservable.  
Given present technology, 
and constraints imposed on the lattice spacing by other observables,
we
conclude that there is little chance to see such an effect in the
atomic spectrum.

\subsubsection{The Energy-Momentum Relation and Cosmic Rays}

\noindent 
Constraints on Lorentz-violating perturbations to the
standard model of electroweak interactions from differences in the
maximal attainable velocity (MAV) of particles (e.g. Ref.~\cite{Coleman:1998ti}),
and on interactions with a non-zero vector field (e.g. Ref.~\cite{Maccione:2009ju}),
have been determined
previously.  
Assuming that each
particle satisfies an energy-momentum relation of the form $E_i^2 =
|{\bf p}_i|^2 c_i^2 + m_i^2 c_i^4$ (along with the conservation of
both energy and momentum in any given process), if $c_\gamma$ exceeds
$c_{e^\pm}$, the process $\gamma\rightarrow e^+ e^-$ becomes possible
for photons with an energy greater than the critical energy $E_{\rm
  crit.} = 2 m_e c_e^2 c_\gamma /\sqrt{ c_\gamma^2-c_{e^\pm}^2}$, and
the observation of high energy primary cosmic photons with
$E_\gamma\lsim 20~{\rm TeV}$ translates into the constraint $c_\gamma
- c_{e^\pm} \lsim 10^{-15}$.  Ref.~\cite{Coleman:1998ti} presents a
series of exceedingly tight constraints on differences between the
speed of light between different particles, with typical sizes of
$\delta c_{ij}\lsim 10^{-21} - 10^{-22}$ for particles of species $i$
and $j$.  At first glance, these constraints~\cite{Gagnon:2004xh}
would appear to also provide tight constraints on the size of the
lattice spacing used in a simulation of the universe.  However, this
is not the case.  As the speed of light for each particle 
in the discretized space-time depends on
its three-momentum, the constraints obtained by Coleman and
Glashow~\cite{Coleman:1998ti} do not directly apply to processes
occurring in a lattice simulation.

\begin{figure}[!t]
\begin{tabular}{cc}
\includegraphics[width=0.7\textwidth]{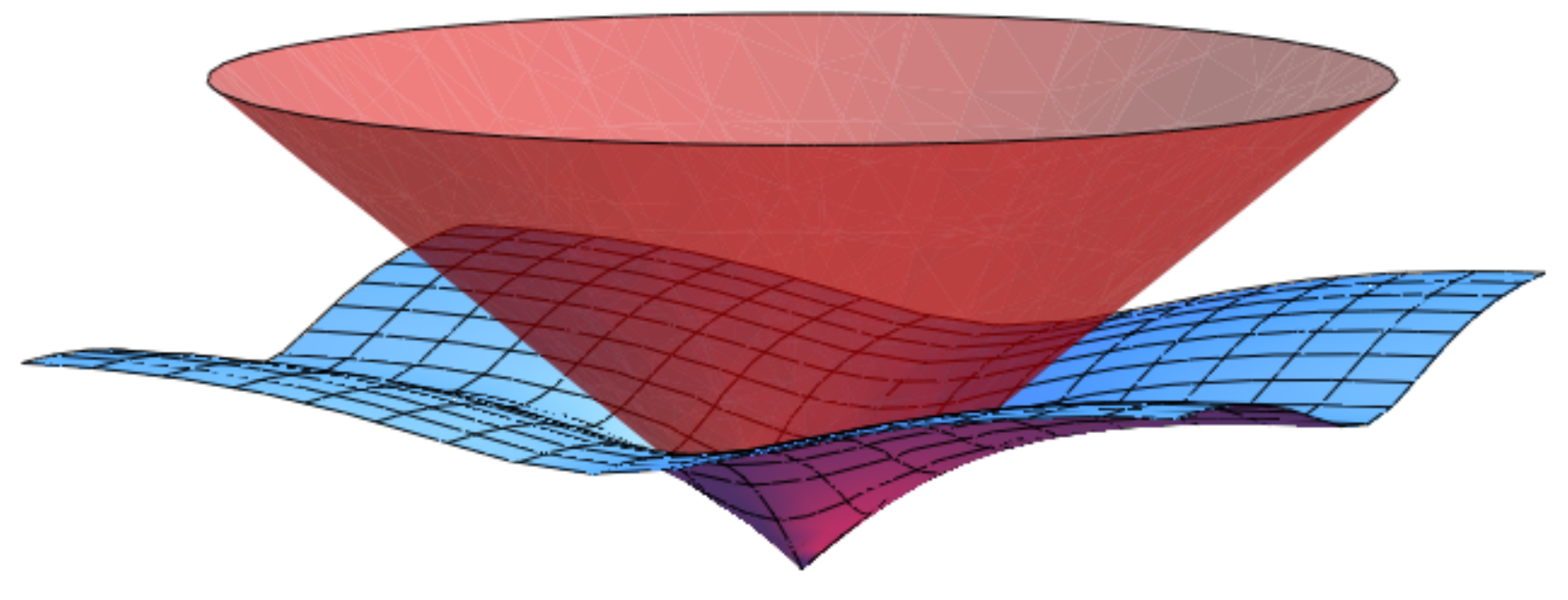}
\end{tabular}
\caption{\label{fig:Epxpy}
The energy surface of a massless, non-interacting 
Wilson fermion with $r=1$  as a function of momentum in the x and y
directions, bounded by $-\pi < b p_{x,y} < \pi$,
for $p_z=0$ is shown in blue. 
The continuum dispersion relation is shown as the red surface.
}
\end{figure}

The dispersion relations satisfied by bosons and Wilson fermions in a
lattice simulation (in
Minkowski space) are 
\begin{eqnarray}
&&
\sinh^2({b E_b\over 2})\ -\ 
\sum_{j=1,2,3} \sin^2({b k_j\over 2})
\ - \ ({b m_b\over 2})^2 \ = \ 0 \ ;
\nonumber\\
&& \ \ \ \ \ \ \ \ \ \ \
E_b 
\ =\ 
\sqrt{|{\bf k}|^2+m_b^2}
\ +\ {\cal O}(b^2) 
\ \ \ ,
\label{dispersionrelationsb}
\end{eqnarray}
and 
\begin{eqnarray}
&& 
\sinh^2\left({b E_f}\right)-
\sum_{j=1,2,3} \sin^2({b k_j}) - 
\left[ 
b m_f + 2 r \left( \sum_{j=1,2,3} \sin^2({b k_j\over 2}) -
  \sinh^2({b E_f\over 2}) \right)\right]^2
\ =\ 0 \ ;
\nonumber\\
&& \ \ \ \  \ \ \ \ \ \ \ \  \ \ \ \ \ \ \ \  \ \ \
E_f 
\ =\ 
\sqrt{|{\bf k}|^2+m_f^2}
\ - \ {r \ b \ m_f^3 \over 2 \sqrt{|{\bf k}|^2+m_f^2}}
\ +\ {\cal O}(b^2)
\ \ \ ,
\label{dispersionrelationsf}
\end{eqnarray}
respectively, where  $r$ is the coefficient of the Wilson term, $E_b$ and
$E_f$ are the energy of a boson and fermion with momentum ${\bf k}$,
respectively. The summations are performed over the components
along the lattice Cartesian axes  corresponding to the x,y, and z spatial
directions.  The implications of these dispersion relations for
neutrino oscillations along one of the lattice axes have been
considered in Ref.~\cite{Motie:2012qj}.  Further, they have been
considered as a possible explanation~\cite{Xue:2011tz} of the (now
retracted) OPERA result suggesting superluminal
neutrinos~\cite{Adam:2011zb}.  The violation of Lorentz invariance
resulting from these dispersion relations is due to the fact that they
have only cubic symmetry and not full rotational symmetry, as shown in
fig.~\ref{fig:Epxpy}.
It is in the limit of small momentum, compared to the inverse lattice
spacing, that the dispersion relations exhibit rotational
invariance.  While for the fundamental particles, the dispersion
relations in 
eq.~(\ref{dispersionrelationsb}) and
eq.~(\ref{dispersionrelationsf}) 
are valid, for composite
particles, such as the proton or pion, the dispersion relations will
be dynamically generated.  
In the present analysis we assume that the
dispersion relations for all particles take the form of those in
eq.~(\ref{dispersionrelationsb}) and
eq.~(\ref{dispersionrelationsf}).  
It is also interesting to note
that the
polarizations of the massless vector fields are not exactly perpendicular to
their direction of propagation for some directions of propagation with respect
to the lattice axes, with longitudinal components present
for non-zero lattice spacings.

Consider the process $p\rightarrow p + \gamma$, which is forbidden in
the vacuum by energy-momentum conservation in special relativity when
the speed of light of the proton and photon are equal, $c_p=c_\gamma$.
Such a process can proceed in-medium when $v_p > c_\gamma$,
corresponding to Cerenkov radiation.  In the situation where the
proton and photon have different MAV's, the absence of this process in
vacuum requires that $|c_p-c_\gamma|\lsim
10^{-23}$~\cite{Coleman:1997xq,Coleman:1998ti}.  In lattice
simulations of the universe, this process could proceed in the vacuum
if there are final state momenta which satisfy energy conservation for
an initial state proton with energy $E_i$ moving in some direction
with respect to the underlying cubic lattice.  Numerically, we find
that there are no final states that satisfy this condition, and
therefore this process is forbidden for all proton
momentum~\footnote{A more complete treatment of this process involves
  using the parton distributions of the proton to relate its energy to
  its momentum~\cite{Gagnon:2004xh}.  For the composite proton, the $p\rightarrow p +
  \gamma$ process becomes kinematically allowed, but with a rate that
  is suppressed by ${\cal O}(\Lambda_{\rm QCD}^8 b^7)$ due to the
  momentum transfer involved, effectively preventing the process from
  occuring.  With momentum transfers of the scale $\sim 1/b$, the
  final states that would be preferred in inclusive decays,
  $p\rightarrow X_h + \gamma$, are kinematically forbidden, with
  invariant masses of $\sim 1/b$.  More refined explorations of this
  and other processes are required.}.  In contrast, the process
$\gamma\rightarrow e^+e^-$, which provides tight constraints on
differences between MAV's~\cite{Coleman:1998ti}, can proceed for very
high energy photons (those with energies comparable to the inverse
lattice spacing) near the edges of the Brillouin zone.  Further, very
high energy $\pi^0$'s are stable against
$\pi^0\rightarrow\gamma\gamma$, as is the related process
$\gamma\rightarrow \pi^0\gamma$.

With the dispersion relation of special relativity, the structure of
the cosmic ray spectrum is greatly impacted by the inelastic
collisions of nucleons with the cosmic microwave background
(CMB)~\cite{Greisen:1966jv,Zatsepin:1966jv}.  Processes such as
$\gamma_{\rm CMB}+N\rightarrow\Delta$ give rise to the predicted
GKZ-cut off scale~\cite{Greisen:1966jv,Zatsepin:1966jv} of $\sim
6\times 10^{20}~{\rm eV}$ in the spectrum of high energy cosmic rays.
Recent experimental observations show a decline in the fluxes starting
around this value~\cite{Abraham:2010mj,Sokolsky:2010kb}, indicating
that the GKZ-cut off (or some other cut off mechanism) is present in
the cosmic ray flux.  For lattice spacings corresponding to an energy
scale comparable to the GKZ cut off, the cosmic ray spectrum will
exhibit significant deviations from isotropy, revealing the cubic
structure of the lattice.  However, for lattice spacings much smaller
than the GKZ cut off scale, the GKZ mechanism cuts off the spectrum,
effectively hiding the underlying lattice structure.  
When the lattice rest frame coincides with the CMB rest frame, 
head-on interactions  between a high energy proton with momentum
$|{\bf p}|$ and a photon of (very-low) energy $\omega$ can proceed through the $\Delta$
resonance when
\begin{eqnarray}
\omega & = & {m_\Delta^2-m_N^2\over 4|{\bf p}|}
\left[ 1 + 
{ \sqrt{\pi}\lattspace^2 |{\bf p}|^2\over 9}
\left( Y_4^0(\theta,\phi)\ +\ 
\sqrt{5\over 14}\left( Y_4^{+4}(\theta,\phi)\ +\  Y_4^{-4}(\theta,\phi)\right)\right)
\right]
\nonumber\\
&&\ -\ {m_\Delta^3-m_N^3\over 4|{\bf p}|} b r
\ +\ ...
\ \ \ ,
\label{eq:gkzkins}
\end{eqnarray}
for $|{\bf p}|\ll 1/\lattspace$, where $\theta$ and $\phi$ are the polar and azimuthal angles of the
particle momenta in the rest frame of the lattice, respectively.
This represents a lower bound for the energy of photons participating in
such a process with arbitrary collision angles.

The lattice spacing itself introduces a cut off to the cosmic ray
spectrum.  
For both the fermions and the bosons, the cut off from the dispersion relation is
$E^{\rm max} \sim 1/ \lattspace$.
Equating this to the GKZ cut off
corresponds to a lattice spacing of $b\sim 10^{-12}~{\rm fm}$, or a
mass scale of $b^{-1}\sim 10^{11}~{\rm GeV}$.  Therefore, the
lattice spacing used in the lattice simulation of the universe must be
$b \lsim 10^{-12}~{\rm fm}$ in order for the GZK cut off to be present
or for the lattice spacing itself to provide the cut off in the cosmic
ray spectrum.  The most striking feature of the scenario in which the
lattice provides the cut off to the cosmic ray spectrum is that the
angular distribution of the highest energy components would exhibit
cubic symmetry in the rest frame of the lattice, 
deviating significantly from isotropy.  For smaller
lattice spacings, the cubic distribution would be less significant,
and the GKZ mechanism would increasingly dominate the high energy
structure.
It may be the case that
more advanced simulations will  be performed with non-cubic
lattices.  The results obtained for cubic lattices indicate that
the symmetries of the non-cubic lattices should  be imprinted, at some level, 
on the high energy cosmic ray spectrum.

\section{Conclusions}
\label{sec:conc}
\noindent 
In this work, we have taken seriously the possibility that our
universe is a numerical simulation.  In particular, we have explored a
number of observables that may reveal the underlying structure of a
simulation performed with a rigid hyper-cubic space-time grid.  This is
motivated by the progress in performing lattice QCD calculations
involving the fundamental fields and interactions of nature in
femto-sized volumes of space-time, and by the simulation hypothesis of
Bostrom~\cite{Bostrom2003}.  A number of elements required for a
simulation of our universe directly from the fundamental laws of
physics have not yet been established, and we have assumed that they will,
in fact, be developed at some point in the future; two important
elements being an algorithm for simulating chiral gauge theories, and
quantum gravity.  It is interesting to note that in the simulation
scenario, the fundamental energy scale defined by the lattice spacing
can be orders of magnitude smaller than the Planck scale, in which
case the conflict between quantum mechanics and gravity should be
absent.

The spectrum of the highest energy cosmic rays provides the most
stringent constraint that we have found on the lattice spacing of a
universe simulation, but precision measurements, particularly the muon
$g-2$, are within a few orders of magnitude of being sensitive to the
chiral symmetry breaking aspects of a simulation employing the
unimproved Wilson lattice action.  
Given the ease with which current
lattice QCD simulations incorporate improvement or employ 
discretizations that preserve chiral symmetry, 
it seems unlikely
that any but the very earliest universe simulations would be
unimproved with respect to the lattice spacing. 
Of course, improvement
in this context masks much of our ability to probe the possibility that our
universe is a simulation, and we have seen that, with the exception of
the modifications to the dispersion relation and the associated maximum values
of energy and momentum, even ${\cal O}(b^2)$
operators in the Symanzik action easily avoid obvious experimental
probes. Nevertheless, assuming that the universe is finite and
therefore the resources of potential simulators are finite, then
a volume containing a simulation will be finite and a lattice spacing
must be non-zero, and therefore in principle there always remains the
possibility for the simulated to discover the simulators.

\subsection*{Acknowledgments}

We would like to thank Eric Adelberger, Blayne Heckel, David Kaplan, 
Kostas Orginos, Sanjay Reddy and Kenneth Roche for
interesting discussions. 
We also thank William Detmold, Thomas Luu  and Ann Nelson for comments on earlier versions
of the manuscript.
SRB was partially supported by the INT during
the program INT-12-2b: Lattice QCD studies of excited resonances and
multi-hadron systems, and by NSF continuing grant PHY1206498. 
In addition, SRB gratefully acknowledges the hospitality of HISKP and the
support of the
Mercator programme of the Deutsche Forschungsgemeinschaft.  
ZD and MJS
were supported in part by the DOE grant DE-FG03-97ER4014.

\bibliography{bibi}
\end{document}